\documentclass[aps,prd,preprint,preprintnumbers,showpacs]{revtex4-1}
\usepackage{graphicx}
\usepackage{amsmath}
\usepackage{caption}
\usepackage[section] {placeins}
\usepackage{slashed}
\usepackage[dvips]{color}
\usepackage{soul}
\usepackage{multirow}
\begin{document}

\title{Negative parity $\Lambda$ and $\Sigma$ resonances coupled to pseudoscalar and vector mesons. }

\author{K.~P.~Khemchandani$^1$\footnote{kanchan@rcnp.osaka-u.ac.jp}}
\author{A.~Mart\'inez~Torres$^2$\footnote{amartine@yukawa.kyoto-u.ac.jp}}
\author{ H.~Nagahiro$^{1, 3}$\footnote{nagahiro@rcnp.osaka-u.ac.jp} }
\author{ A.~Hosaka$^1$\footnote{hosaka@rcnp.osaka-u.ac.jp}}
\preprint{YITP-12-23}

 \affiliation{
$^1$ Research Center for Nuclear Physics (RCNP), Mihogaoka 10-1, Ibaraki 567-0047, Japan.\\
$^2$ Yukawa Institute for Theoretical Physics, Kyoto University, Kyoto 606-8502, Japan.\\
$^3$ Department of Physics, Nara Women's University,  Nara 630-8506, Japan.
}

\date{\today}

\begin{abstract}
An exploration of the dynamically generated $\Lambda$ and $\Sigma$ resonances has been made in the vector meson-baryon (VB) systems coupled to pseudoscalar meson-baryon (PB) channels.  The VB interactions are obtained from the Lagrangians written within the hidden local symmetry which, in addition to Yukawa-type vector-baryon-baryon (VBB) vertices, naturally gives rise to a contact interaction. Using the VBB vertices, we calculate the $t$-, $s$- and $u$-channel diagrams considering the octet baryon exchange in the latter two cases. For the PB  channels we rely on the  the Weingberg-Tomozawa interactions and calculate the PB $\leftrightarrow$ VB transition amplitudes extending the Kroll-Ruderman theorem by replacing the photon by the vector mesons. 
In addition to the low-lying $\Lambda(1405)$, $\Lambda(1670)$ found in previous works, we find resonances which coincide well with: $\Lambda(2000)$, $\Sigma(1750) S_{11}$, $\Sigma(1940) D_{13}$, $\Sigma(2000) S_{11}$ and predict three $\Lambda^*$'s with $J^\pi = 3/2^-$ and masses: $1754 - i2$ MeV, $1862 - i33$ MeV, $2139 - i14$ MeV together with a broad $1/2^-$  $\Sigma^*$ with mass $\sim$ 1430 MeV. We find that most of  these next-to-low-lying states couple weakly with the pseudoscalar mesons, which implies that it is important to study the reactions with VB final states in order to obtain a reliable information on the properties of  such resonances.

\end{abstract}

\maketitle

\section{Introduction}

Understanding  meson-baryon interaction is an important aspect of exploring the strong interaction at low and intermediate energies. A lot of work has been done in this direction, especially in case of  the low energy interaction of light pseudoscalar mesons and baryons, where several resonances in the baryon spectrum arise due to this interaction \cite{wyld,kaiser,baryon1,eulogiopb,baryon2,jidolambda,oller2006,oller2007,oka,hyodo12,hyodoreport}. These states can be understood as weakly bound meson-baryon systems, whose existence originates directly from the characteristics of the theory of quantum chromodynamics (QCD), which leads to the generation of a quark-antiquark pair when the distance between the quarks increases. In other words, this leads to the decay of a heavier hadron to two (or more) lighter ones. Such a decaying heavier hadron can be understood as a quasi-bound system  of two (or more)  lighter hadrons (for baryon resonances found in two, three (or more) meson-baryon systems see, for example, Refs.~\cite{3b1,3b2,3b3}). In case of the pseudoscalar meson-baryon system, the low energy interaction has been found to be governed by the spontaneous breaking of the chiral symmetry of   QCD and is well explained in terms of the low energy theorems. Studies of relevant systems, accomplished with the effective field theories based on  chiral symmetry and its spontaneous breaking have been very successful in explaining the properties of some of the baryon resonances. 

Recently, some attention is being  paid to the vector meson-baryon (VB) interactions as well \cite{lutz,juelich,michael,pedro,eulogiovb,sourav,us,javi}. The difficulty in this case comes from the fact that vector mesons are not light enough for the low energy theorems to be applicable. Considering this, we have recently studied the vector meson-baryon interaction within a formalism based on the hidden local symmetry (HLS), which accommodates vector mesons consistently with the chiral symmetry. In this work we have calculated the $t$-, $s$- and $u$-channel exchange diagrams together with a contact interaction which arises from the part of the HLS Lagrangian related to the anomalous magnetic moment of the baryons. It was shown in Ref.~\cite{us} that this latter contact interaction is demanded by the gauge invariance of the HLS Lagrangian containing the term related to the anomalous magnetic moment of the baryons. The contribution from all the diagrams, except the $s$-channel exchange, considered in Ref.~\cite{us} was shown to be important for the systems with total strangeness zero. The purpose of the present work is to study the strangeness $-1$ VB systems.

Further, in an independent study, we checked the effect of coupling vector mesons  to the resonances generated in the PB systems in Ref.~\cite{us2}.  Some earlier works have shown that certain baryon resonances couple strongly to both  pseudoscalar and vector mesons   \cite{lutz,juelich,michael}. In Ref.~\cite{us2} we made a coupled channel calculations taking both PB and VB channels into account for total strangeness $-1$ but since the focus was on  the low-lying resonances, we  calculated the PB and VB diagonal amplitudes by considering only the $t$-channel diagrams which simplified the formalism.  The transition amplitudes between the two channels were obtained by extending the Kroll-Ruderman theorem by replacing the photon by a vector meson under the vector meson dominance notion. Our calculations done in this formalism lead to very interesting results. We found that the low-lying $\Lambda$ and $\Sigma$ resonances couple strongly to the closed, heavy mass, VB channels. These results do not imply  that the wave functions of the low-lying strangeness $-1$ resonances consists of important VB components because the heavy mass of  the VB channels suppresses it. However, the findings of Ref.~\cite{us2} are very important for building models to study reactions producing such resonances. Although the motivation for the study reported in  Ref.~\cite{us2} was to study low-lying resonances, we tested if the heavier resonances became wider by coupling them to PB channels but an interpretation of  the higher energy resonances could not be made  conclusively since the $t$-channel diagrams  were considered as the only source of VB interactions.
Thus, it remains to study the effect of coupling pseudoscalar mesons to VB resonances, which is appropriate to make in the present work and use the wisdom gained from Refs.~\cite{us, us2} together.

The paper is organized as follows: in the next section we obtain the interaction between the vector mesons and baryons with total strangeness $-1$ and discuss the transition amplitudes between VB and PB systems. We show the amplitudes obtained on the real axis and discuss the related poles found in the complex plane. Finally, we give a summary of the resonances found in the present work and their relation to the known resonances..

\section{Vector Meson-Baryon interaction.}\label{intI}

The foundation of our formalism  lays on the gauge invariant vector meson-baryon (VB)  Lagrangian written as 
\begin{equation}
\mathcal{L_{\rm VB}} =   \bar{\psi}  i \slashed D  \psi, \label{one}
\end{equation}
which has been obtained  through the minimal substitution 
\begin{equation}
\partial_\mu \longrightarrow D_\mu = \partial_\mu + i g \rho_\mu (x),
\end{equation}
and by  requiring that the nucleon fields ($\psi$) transform under the hidden local symmetry (HLS)  as 
$\psi \rightarrow h(x) \psi$, where $h (x)$ is an element of the HLS.  We set the sign convention such that, analogously  to the quantum electrodynamics, 
a positive sign for the amplitude $ V = - \mathcal{L} = g \bar{\psi} \gamma^\mu \rho_\mu \psi$  (for $g > 0$ ) arises for a positively charged  field. Further, keeping in mind the importance of the reproduction of the anomalous 
magnetic moment of the baryons, we extend the interaction term of Eq.~(\ref{one}) (in SU(2)) to
\begin{equation}
\mathcal{L}_{\rho N } = - g \bar{\psi} \left\{ \gamma_\mu \rho^\mu + \frac{\kappa_\rho}{4M} \sigma_{\mu\nu} \rho^{\mu\nu} \right\} \psi, \label{rhoNL}
\end{equation}
with 
\begin{equation}
\rho^{\mu\nu} = \partial_\mu \rho_\nu -  \partial_\nu \rho_\mu +  i g \left[ \rho_\mu, \rho_\nu \right]. 
\end{equation}
It was  discussed in detail in Ref.~\cite{us}, that the hidden gauge invariance of the Lagrangian given by Eq.~(\ref{rhoNL}) requires the invariance of the new term: $ \bar{\psi} h^\dagger(x)  \sigma_{\mu\nu}  \rho^{\mu\nu} h (x) \psi$,
which can be accomplished only when the commutator part of the tensor field $\rho^{\mu\nu}$ is taken into account. Consequently, one is left with a contact VB interaction which should be essentially considered.  Further, this interaction was  found to be  large
in Ref.~\cite{us}. Additionally, the analytical calculations of the amplitudes corresponding to the exchanges of the octet baryons in the $s$- and $u$-channels, at the leading order, were also found to be comparable in order of magnitude to those obtained from the $t$-channel and the contact interactions. 
All these findings motivated the SU(3) generalization of the Lagrangian of Eq.~(\ref{rhoNL}) to
\begin{eqnarray} \nonumber
\mathcal{L}_{\rm VBB}&=& -g \biggl\{ \langle \bar{B} \gamma_\mu \left[ V_8^\mu, B \right] \rangle + \frac{1}{4 M} \Bigl( F \langle \bar{B} \sigma_{\mu\nu} \left[ \partial^{\mu} V_8^\nu - \partial^{\nu} V_8^\mu, B \right] \rangle \Bigr.
\biggr.   \\ \nonumber
&+&\Bigl.  D \langle \bar{B} \sigma_{\mu\nu} \left\{ \partial^{\mu} V_8^\nu - \partial^{\nu} V_8^\mu, B \right\} \rangle\Bigr)
+ \langle \bar{B} \gamma_\mu B \rangle  \langle  V_0^\mu \rangle  \\
&+&\left.  \frac{ C_0}{4 M}  \langle \bar{B} \sigma_{\mu\nu}  V_0^{\mu\nu} B  \rangle\right\},\label{yukawaL}
\end{eqnarray}
for the Yukawa type vertices and 
\begin{equation}
\mathcal{L}_{VVBB} = - \frac{g}{4 M} \Bigl\{ F \langle \bar{B} \sigma_{\mu\nu} \left[ ig \left[V_8^\mu, V_8^\nu \right], B \right] \rangle  + D \langle \bar{B} \sigma_{\mu\nu} \left\{  ig \left[V_8^\mu, V_8^\nu \right], B \right\} \rangle \Bigr\}. \label{contactl}
\end{equation}
for the contact interactions, where the subscripts $8$ and $0$ denote the octet and singlet vector fields, respectively, which have been obtained by considering the $\omega-\phi$ mixing into account. The constants $D$ and $F$ in Eq.~(\ref{yukawaL}) are fixed to the values $2.4$ and  $0.82$, respectively, as in  Ref.~\cite{jido}, where a good reproduction of the  magnetic moments of the baryons was obtained. Further,  the constant for the singlet vector meson-baryon interaction $C_0$ is taken as $3F - D$. In this way, we have the anomalous magnetic couplings of the vector meson-baryon-baryon vertices consistent with their known
values: $\kappa_\rho \simeq 3.2, \kappa_\omega \simeq \kappa_\phi \simeq 0$. 

The Lagrangian given by Eq.~(\ref{contactl}) leads to the  amplitude of the vector meson-baryon interaction 
\begin{equation}
V^{CT}_{ij} = i C^{CT}_{ij} \frac{g^2}{ \sqrt{4 M_i M_j} }\vec{\sigma}\cdot\vec{\epsilon_2} \times \vec{\epsilon_1},\label{contactv}
\end{equation}
where $M_i (M_j)$ refers to the mass of the baryon in the initial (final) state and  $\vec{\epsilon}_1 (\vec{\epsilon}_2)$, here and throughout this article, denotes the polarization vector of the vector meson in the initial (final) state. Clearly, this 
interaction possesses a $\vec{s} \cdot \vec{S}$ structure, where $\vec{s}$ and $\vec{S}$ represent the spin  1/2 and spin $1$ operators, respectively.  
The constants $C^{CT}_{ij}$ in Eq.~(\ref{contactv}) for the VB systems with strangeness $-1$ and isospin $0$ and $1$ configurations are given in Tables~\ref{con0}, and~\ref{con1}, respectively.
\begin{table}[h!]
\caption{Coefficients for the contact term potential for VB systems in the $S=-1$ and $I=0$ configuration.} \label{con0}
\begin{ruledtabular}
\begin{tabular}{cccccc}
                                    &$\bar{K}^* N$                                                  & $\omega \Lambda$                                            & $\rho  \Sigma$                                                  &$\phi \Lambda$                      & $K^* \Xi$                                   \\ \hline
$\bar{K}^* N$           &$\frac{D + 3 F}{2}$                                          & $\frac{D + 3 F}{2 \sqrt{6}}$                                & $\sqrt{\frac{3}{2}}\left(\frac{D - F}{2}\right)$  &$-\frac{D + 3 F}{2 \sqrt{3}}$   & $0$ \\
$\omega \Lambda$&                                                                           & $0$                                                                        & $0$                                                                      &$0$                                           & $\frac{D - 3 F}{2 \sqrt{6}}$ \\
$\rho \Sigma$          &                                                                           &                                                                                & $2 F$                                                                  &$ 0$                                           & $ \sqrt{\frac{3}{2}} \left(\frac{D + F}{2}\right)$ \\
$\phi \Lambda$       &                                                                           &                                                                                &                                                                             &$0$                                            &$-\frac{D - 3 F}{2 \sqrt{3}}$\\
$K^* \Xi$                   &                                                                          &                                                                                &                                                                             &                                                    & $-\frac{D - 3 F}{2}$ 
\end{tabular}
\end{ruledtabular}
\end{table}

\begin{table}[htbp]
\caption[]{Coefficients for the contact term potential for VB systems in  the $S=-1$ and $I=1$ configuration.} \label{con1}
\begin{ruledtabular}
\begin{tabular}{c|cccccc}
                                 &$\bar{K}^* N$                        &$\rho \Lambda$                    & $\rho \Sigma$                       &$\omega \Sigma$                 &$K^* \Xi$                               & $\phi \Sigma$  \\ \hline
$\bar{K}^* N$        &$-\frac{D - F}{2}$                   &$-\frac{D + 3 F}{2 \sqrt{6}}$ & $-\frac{F - D}{2} $                 &$\frac{D - F}{2 \sqrt{2}}$       &$0$                                        & $-\frac{D - F}{2}$ \\
$\rho \Lambda$    &                                                 &$0$                                          & $-\sqrt{\frac{2}{3}} D$          &$0$                                          &$\frac{D - 3 F}{2 \sqrt{6}}$  & $0$ \\
$\rho \Sigma$       &                                                 &                                                 & $F$                                         &$0$                                          &$\frac{D + F}{2}$                  & $0$ \\ 
$\omega \Sigma$&                                                 &                                                 &                                                 & $0$                                         & $-\frac{D + F}{2 \sqrt{2}}$   & $0$ \\
$K^* \Xi$               &                                                 &                                                 &                                                 & $0$                                         &$\frac{D + F}{2}$                   &$\frac{D + F}{2}$\\
$\phi \Sigma$       &                                                 &                                                 &                                                 &                                                 &                                                & $0$
\end{tabular}
\end{ruledtabular}
\end{table}

The amplitudes for the $s$-, $u$- and $t$-channel exchanges can be obtained using Eq.~(\ref{yukawaL}) and the three-vector Lagrangian contained in the kinetic herm of the vector field;
\begin{equation}
\mathcal{L}_{3V} \in - \frac{1}{2} \langle V^{\mu\nu} V_{\mu\nu} \rangle.
\end{equation}
For the sake of completeness, we would like to mention at this point that we use the standard approach of obtaining the  $s$-, $u$- and $t$-channel amplitudes within the non-relativistic limit, which is suitable for studies of hadronic systems at near-threshold energies. Such diagrams, thus, 
 effectively give rise to point interactions, which have the following general form:
 \begin{eqnarray} 
V^{t}_{ij} &=& - C^{t}_{ij} \frac{1}{4 f_\pi^2} (K_1^0 + K_2^0) \vec{\epsilon_1}.\vec{\epsilon_2}, \label{vwt}\\
V^{u}_{ij}&=&C^u_{ij} \left(-\frac{g^2}{m-2 M}\right) \vec{\epsilon}_1\cdot \vec{\sigma}\,\, \vec{\epsilon}_2\cdot \vec{\sigma },\label{vu} \\
V^{s}_{ij} &=&C^s_{ij} \left(\frac{g^2}{m+2 M}\right) \vec{\epsilon}_2\cdot \vec{\sigma} \,\,\vec{\epsilon}_1\cdot \vec{\sigma }.\label{vs}
\end{eqnarray}
The coefficients $C^{t}_{ij}, C^{u}_{ij}, C^{s}_{ij}$ have been listed in Ref.~\cite{us} for  the meson-baryon systems with total strangeness $0$.
 We give the $C_{ij}$ coefficients of Eq.~(\ref{vwt}-\ref{vs}) for the  strangeness $-1$ systems with isospin $0$ as well as $1$  in: Tables~\ref{u_iso0},  \ref{u_iso1} for the $u$-channel and  Tables~\ref{s_iso0}, \ref{s_iso1} for the $s$-channel, respectively. The $C^t_{ij}$'s (for the $t$-channel) can be found  in Refs.~\cite{eulogiopb, us2}.

{\squeezetable
\begin{turnpage}
\begin{table}[htbp!]
\caption[]{Coefficients for the U-channel potential for $S=-1$ and $I=0$.} \label{u_iso0}
\begin{ruledtabular}
\begin{tabular}{c|ccccc}
  & $\bar{K}^*N$                   &$\omega \Lambda$   & $\rho  \Sigma$ & $\phi \Lambda$ & $K^* \Xi$ \\
  \hline\\
$\bar{K}^*N$              &$0$ & $\frac{-36 M^2-36 F m M+\left(D^2-9 F^2\right) m^2}{8 \sqrt{6} M^2}$  &$ \frac{\sqrt{\frac{3}{2}} \left(4 M^2+4 F m M+\left(F^2-D^2\right) m^2\right)}{8 M^2}$      & $0$
                                                & $\frac{36 M^2+36 F m M+\left(9 F^2-5 D^2\right) m^2}{24 M^2}$  \\                                 
$\omega \Lambda$  &        & $\frac{\left(6 M+\left(3 F-2 D\right) m\right)^2}{36 M^2}$                           & $-\frac{D m \left(2 M+F m\right)}{4 M^2}$                                                                                  &$\frac{\left(6 M+\left(3 F-2 D\right) m\right) \left(6 M+\left(D+3 F\right) m\right)}{36 \sqrt{2} M^2}$ & $-\frac{\left(6 M-\left(D-3 F\right) m\right) \left(2 M+\left(F-D\right) m\right)}{8 \sqrt{6} M^2}$\\
$\rho \Sigma$            &        &                                                                                                                              & $\frac{-24 M^2-24 F m M+\left(D^2-6 F^2\right) m^2}{12 M^2}$                                          & $\frac{D m \left(\left(D-F\right) m-2 M\right)}{4 \sqrt{2} M^2}$ 
                                                & $-\frac{\sqrt{\frac{3}{2}} \left(4 M^2+4 F m M+\left(F^2-D^2\right) m^2\right)}{8 M^2}$ \\
$\phi \Lambda$         &        &                                                                                                                              &                                                                                                                                                          & $\frac{\left(6 M+\left(D+3 F\right) m\right)^2}{72 M^2}$                                     
                                                & $-\frac{\left(6 M-\left(D-3 F\right) m\right) \left(2 M+F m\right)}{4 \sqrt{3} M^2}$\\
$K^* \Xi$                     &        &                                                                                                                              &                                                                                                                                                          &
                                                & $0$ 
 \end{tabular}
\end{ruledtabular}
\end{table}
\end{turnpage}}
  It is important to recall that  the contribution of the $s$- and $u$-channel diagrams  in the present study of s-wave (near-threshold) meson-baryon interaction comes only from the terms involving   the negative energy solution of the Dirac equation for the baryon propagator.  Since such diagrams require large momentum transfers at the non-relativistic energies, we include a form factor in the calculations of these diagrams which is written following Refs.~\cite{Naam,formfactormart,mosel1,mosel2} as:
\begin{equation}
F(\Lambda, x) = \frac{\Lambda^4}{\Lambda^4 + \left( x^2 - M_x^2\right)^2},\label{formfactor}
\end{equation}  
where $x$ is the Mandelstam variable under consideration ($s$ or $u$), $M_x$ is the  mass of the baryon exchanged in such diagrams
and $\Lambda$ is a parameter which we fix as 650~MeV since it  corresponds to a reasonable average size of the hadrons ($\sim$ 0.6 fm).{\squeezetable
\begin{turnpage}
\begin{table}[htbp!]
\caption[]{Coefficients for the U-channel potential for $S=-1$ and $I=1$.} \label{u_iso1}
\begin{ruledtabular}    
\begin{tabular}{ccccccc}
  & $\bar{K}^*N$     & $\rho \Lambda$ & $\rho \Sigma$   & $\omega \Sigma$ &   $K^* \Xi$ & $\phi \Sigma$  \\
  \hline\\
$\bar{K}^* N$         &$0$                                                                                                               &$\frac{\left(2 M+\left(D+F\right) m\right) \left(6 M+\left(D+3 F\right) m\right)}{8 \sqrt{6} M^2}$&$\frac{-4 M^2-4 F m M+\left(D^2-F^2\right) m^2}{8 M^2}$  
 &$\frac{\left(6 M-\left(D-3 F\right) m\right) \left(2 M+\left(F-D\right) m\right)}{8 \sqrt{2} M^2}$& $-\frac{12 M^2+12 F m M+\left(D^2+3 F^2\right) m^2}{24 M^2}$                                       & $0$ \\
$\rho \Lambda$     &                                                                                                                      & $\frac{D^2 m^2}{12 M^2}$                                                                                                                & $\frac{D m \left(2 M+F m\right)}{2 \sqrt{6} M^2}$  
                                 & $-\frac{D m \left(\left(2 D-3 F\right) m-6 M\right)}{12 \sqrt{3} M^2}$  &$\frac{\left(6 M-\left(D-3 F\right) m\right) \left(2 M+\left(F-D\right) m\right)}{8 \sqrt{6} M^2}$  &$\frac{D m \left(6 M+\left(D+3 F\right) m\right)}{12 \sqrt{6} M^2}$ \\
$ \rho  \Sigma$      &                                                                                                                       &                                                                                                                                                               &$\frac{12 M^2+12 F m M-\left(D^2-3 F^2\right) m^2}{12 M^2}$ 
                                 & $-\frac{\left(2 M+F m\right)^2}{2 \sqrt{2} M^2}$                                     &$\frac{4 M^2+4 F m M+\left(F^2-D^2\right) m^2}{8 M^2}$                                                          &$-\frac{\left(2 M+F m\right) \left(2 M+\left(F-D\right) m\right)}{4 M^2}$ \\
$\omega \Sigma$ &                                                                                                                        &                                                                                                                                                               &
                                 & $\frac{\left(2 M+F m\right)^2}{4 M^2}$                                                    &$-\frac{\left(2 M+\left(F-D\right) m\right) \left(2 M+\left(D+F\right) m\right)}{8 \sqrt{2} M^2}$  &$\frac{\left(2 M+F m\right) \left(2 M+\left(F-D\right) m\right)}{4 \sqrt{2} M^2}$ \\
$K^* \Xi$                &                                                                                                                        &                                                                                                                                                                & 
                                 &                                                                                                                        &$0$                                                                                                                                                         &$-\frac{\left(2 M+F m\right) \left(2 M+\left(D+F\right) m\right)}{4 M^2}$ \\
$\phi \Sigma$        &                                                                                                                        &                                                                                                                                                                &                              
                                 &                                                                                                                        &                                                                                                                                                                &$\frac{\left(2 M+\left(F-D\right) m\right)^2}{8 M^2}$          
\end{tabular}
\end{ruledtabular}
\end{table}
\end{turnpage}}
 {\squeezetable
\begin{turnpage}
\begin{table}[htbp!]
\caption[]{Coefficients for the S-channel potential for $S=-1$ and $I=0$.} \label{s_iso0}
\begin{ruledtabular}
\begin{tabular}{cccccc}
  &$\bar{K}^* N$ & $\omega \Lambda$ & $\rho  \Sigma$  & $\phi \Lambda$ & $K^* \Xi$ \\
  \hline\\
$\bar{K}^* N$              & $\frac{\left(\left(D+3 F\right) m-6 M\right)^2}{24 M^2}$                   & $\frac{\left(6 M+\left(2 D-3 F\right) m\right) \left(\left(D+3 F\right) m - 6 M\right)}{12 \sqrt{6} M^2}$   & $\frac{D m \left(\left(D+3 F\right) m-6 M\right)}{4 \sqrt{6} M^2}$                             
                                       & $-\frac{\left(\left(D+3 F\right) m-6 M\right)^2}{24 \sqrt{3} M^2}$    & $-\frac{-36 M^2+36 F m M+\left(D^2-9 F^2\right) m^2}{24 M^2}$\\
$\omega \Lambda$    &                                                                                                                  & $\frac{\left(6 M+\left(2 D-3 F\right) m\right)^2}{36 M^2}$                                                                          & $\frac{D m \left(6 M+\left(2 D-3 F\right) m\right)}{12 M^2}$                          
                                       & $\frac{\left(6 M+\left(2 D-3 F\right) m\right) \left(6 M-\left(D+3 F\right) m\right)}{36 \sqrt{2} M^2}$       & $-\frac{\left(6 M+\left(D-3 F\right) m\right) \left(6 M+\left(2 D-3 F\right) m\right)}{12 \sqrt{6} M^2}$\\
 $\rho  \Sigma$            &                                                                                                                   &                                                                                                                                                                             & $\frac{D^2 m^2}{4 M^2}$                
                                       & $-\frac{D m \left(\left(D+3 F\right) m-6 M\right)}{12 \sqrt{2} M^2}$ & $-\frac{D m \left(6 M+\left(D-3 F\right) m\right)}{4 \sqrt{6} M^2}$\\
 $\phi \Lambda$         &                                                                                                                    &                                                                                                                                                                             &
                                      & $\frac{\left(\left(D+3 F\right) m-6 M\right)^2}{72 M^2}$                  & $\frac{-36 M^2+36 F m M+\left(D^2-9 F^2\right) m^2}{24 \sqrt{3} M^2}$ \\
 $K^* \Xi$                    &                                                                                                                     &                                                                                                                                                                             &     
                                       &                                                                                                                     & $\frac{\left(6 M+\left(D-3 F\right) m\right)^2}{24 M^2}$ 
 \end{tabular}
\end{ruledtabular}
\end{table}
\end{turnpage}}
 {\squeezetable
\begin{turnpage}
\begin{table}[htbp]
\caption[]{Coefficients for the S-channel potential for $S=-1$ and $I=1$.} \label{s_iso1}
\begin{ruledtabular}     
\begin{tabular}{ccccccc}
                                       & $\bar{K}^* N$         &$\rho \Lambda$                    & $\rho \Sigma$                       &$\omega \Sigma$                 &$K^* \Xi$                               & $\phi \Sigma$ \\
  \hline\\
$\bar{K}^* N$              & $\frac{1}{2} \left[\frac{\left(D-F\right) m}{2 M}+1\right]^2$                                   & $-\frac{D m \left(2 M+\left(D-F\right) m\right)}{4 \sqrt{6} M^2}$                             & $-\frac{\left(2 M+\left(D-F\right) m\right) \left(2 M-F m\right)}{4 M^2}$ 
                                       & $\frac{\left(2 M+\left(D-F\right) m\right) \left(2 M-F m\right)}{4 \sqrt{2} M^2}$ & $\frac{-4 M^2+4 F m M+\left(D^2-F^2\right) m^2}{8 M^2}$                                   & $\frac{\left(2 M+\left(D-F\right) m\right)^2}{8 M^2}$ \\
$\rho \Lambda$          &                                                                                                                                      & $\frac{D^2 m^2}{12 M^2}$                                                                                          &  $\frac{D m \left(2 M-F m\right)}{2 \sqrt{6} M^2}$
                                       & $\frac{D m \left(F m-2 M\right)}{4 \sqrt{3} M^2}$                                                  & $-\frac{D m \left(\left(D+F\right) m-2 M\right)}{4 \sqrt{6} M^2}$                             & $-\frac{D m \left(2 M+\left(D-F\right) m\right)}{4 \sqrt{6} M^2}$ \\
 $\rho \Sigma$             &                                                                                                                                      &                                                                                                                                          & $\frac{\left(F m-2 M\right)^2}{2 M^2}$
                                       & $-\frac{\left(F m-2 M\right)^2}{2 \sqrt{2} M^2}$                                                    & $\frac{\left(2 M-F m\right) \left(2 M-\left(D+F\right) m\right)}{4 M^2}$                   & $-\frac{\left(2 M+\left(D-F\right) m\right) \left(2 M-F m\right)}{4 M^2}$ \\
 $\omega \Sigma$      &                                                                                                                                      &                                                                                                                                          &
                                       & $\frac{\left(F m-2 M\right)^2}{4 M^2}$                                                                   & $-\frac{\left(2 M-F m\right) \left(2 M-\left(D+F\right) m\right)}{4 \sqrt{2} M^2}$    & $\frac{\left(2 M+\left(D-F\right) m\right) \left(2 M-F m\right)}{4 \sqrt{2} M^2}$ \\
 $K^* \Xi$                     &                                                                                                                                      &                                                                                                                                          &
                                       &                                                                                                                                      & $\frac{1}{2} \left(1-\frac{\left(D+F\right) m}{2 M}\right)^2$                                      & $\frac{-4 M^2+4 F m M+\left(D^2-F^2\right) m^2}{8 M^2}$ \\
 $\phi \Sigma$             &                                                                                                                                      &                                                                                                                                          &
                                       &                                                                                                                                      &                                                                                                                                          & $\frac{\left(2 M+\left(D-F\right) m\right)^2}{8 M^2}$
\end{tabular}
\end{ruledtabular}
\end{table}
\end{turnpage}
}
\section{Coupling VB and  PB systems.} \label{intII}
The motivation of our present work is twofold: one comes from the findings of Ref.~\cite{us}, where a detailed study of the vector meson-baryon interaction was done for the total strangeness zero systems. It is natural to explore the strangeness $-1$ systems  next. Further, a formalism was developed in Ref.~\cite{us2} to couple the pseudoscalar and vector mesons to baryon resonances, and a study of  low-lying strangeness $-1$ baryon resonances was made. The choice of the low energy range and strangeness $-1$ systems was made in the latter work recalling that it is in this sector where those states exist which can be considered as the best candidates acquiring the nature of dynamically generated resonances. The work presented in Ref.~\cite{us2} relied on the diagonal interactions obtained from the $t$-channel exchange, which is suitable for studying low-lying resonances and which simplified the calculations.  Within the latter simple formalism, we tested the  energy region of VB resonances too to test if their widths  increased by coupling them to the PB channels and we found that the effect of the PB-VB coupling was not limited to widening of poles. It lead to moving of  the poles in the complex plane, which turned them to either virtual states or shadow poles.  It is, thus,  important to study strangeness $-1$ VB  systems coupled with those of PB with the VB interactions obtained in the previous section. 

 For this, guided by  the Weinberg-Tomozawa theorem, we continue to use the PB $\rightarrow$ PB amplitudes calculated with the $t$-channel diagrams.   Further,  the PB $\leftrightarrow$ VB amplitudes are also used from Ref.~\cite{us2}, which were obtained consistently by starting with the non-linear sigma model and 
 by using the Kroll-Ruderman (KR) theorem to write the Lagrangian for the $\gamma N \rightarrow \pi N$ process and by replacing the $\gamma$ by a vector meson via the  notion of the vector meson dominance. This procedure, which required the introduction of  a vector meson field as a gauge boson of the hidden local symmetry, lead to  the following Lagrangian for the flavor SU(3) case
\begin{eqnarray}
\mathcal{L}_{\rm PBVB} = \frac{-i g_{KR}}{2 f_\pi} \left ( \tilde{F} \langle \bar{B} \gamma_\mu \gamma_5 \left[ \left[ P, V^\mu \right], B \right] \rangle + 
\tilde{D} \langle \bar{B} \gamma_\mu \gamma_5 \left\{ \left[ P, V^\mu \right], B \right\}  \rangle \right), \label{pbvb}
\end{eqnarray}
where the trace $\langle ... \rangle$ has to be calculated in the flavor space and $\tilde{F} = 0.46$, $\tilde{D}=0.8$ such that  $\tilde{F} + \tilde{D} \simeq  g_A = 1.26$. The ratio 
$\tilde{D}/(\tilde{F} + \tilde{D}) \sim 0.63$ here is close to the quark model value of 0.6, and the empirical values of $\tilde{F}$ and $\tilde{D}$ can be found, for example, in Ref.~\cite{Yamanishi:2007zza}.  The coupling between the PB and VB channels, $g_{KR} = 6$, in Eq.~(\ref{pbvb}), has been obtained using the  Kawarabayashi-Suzuki-Riazuddin-Fayazuddin relation \cite{ksrf}: $g_{KR} = m_\rho/\left(\sqrt{2} f_\pi \right) \sim 6$.
Using Eq.~(\ref{pbvb}), the amplitudes for  the processes involving all the strangeness $-1$ PB-VB systems  have been obtained and are given in Ref.~\cite{us2}, which we use in the present work also.

\section{Results and discussions} 
In the previous sections we have given  the information required  to calculate the lowest order amplitudes for different channels coupled to strangeness $-1$ involving pseudoscalar and vector mesons. These amplitudes have been used as the kernels to solve  the
Bethe-Salpeter equations
\begin{equation}
T = V + VGT,\label{bs}
\end{equation}
where the loops are calculated using a Gaussian cut-off 
 \begin{equation}
 G = \int\limits_0^\infty \frac{d^3q}{\left(2 \pi \right)^3} \frac{1}{2 E_1 \left(\vec{q}, m\right)} \frac{2 M}{2 E_2 \left(\vec{q}, M\right)} \frac{ e^{-\left(q^2 - q_{on}^2 \right)/\Lambda^2}}{E - E_1 \left(\vec{q}, m\right) - E_2 \left(\vec{q}, M\right)}, \label{loopcutoff}
 \end{equation} 
with the same cut-off values as those used in Ref.~\cite{us2}: for the PB channels $\Lambda_{\rm PB}=$ 750 MeV and for VB channels $\Lambda_{\rm VB}=$ 545 MeV.  Finally, we would like to add that the loops of the channels involving mesons with large widths ($\rho, K^*$) have been calculated by making a convolution  over the varied mass of these mesons \cite{eulogiovb,us,us2}.

\subsection{ $\Lambda^*$ and $\Sigma^*$ in VB systems}
\begin{figure}[ht!]
\includegraphics[width= 0.38\textwidth,height=18cm]{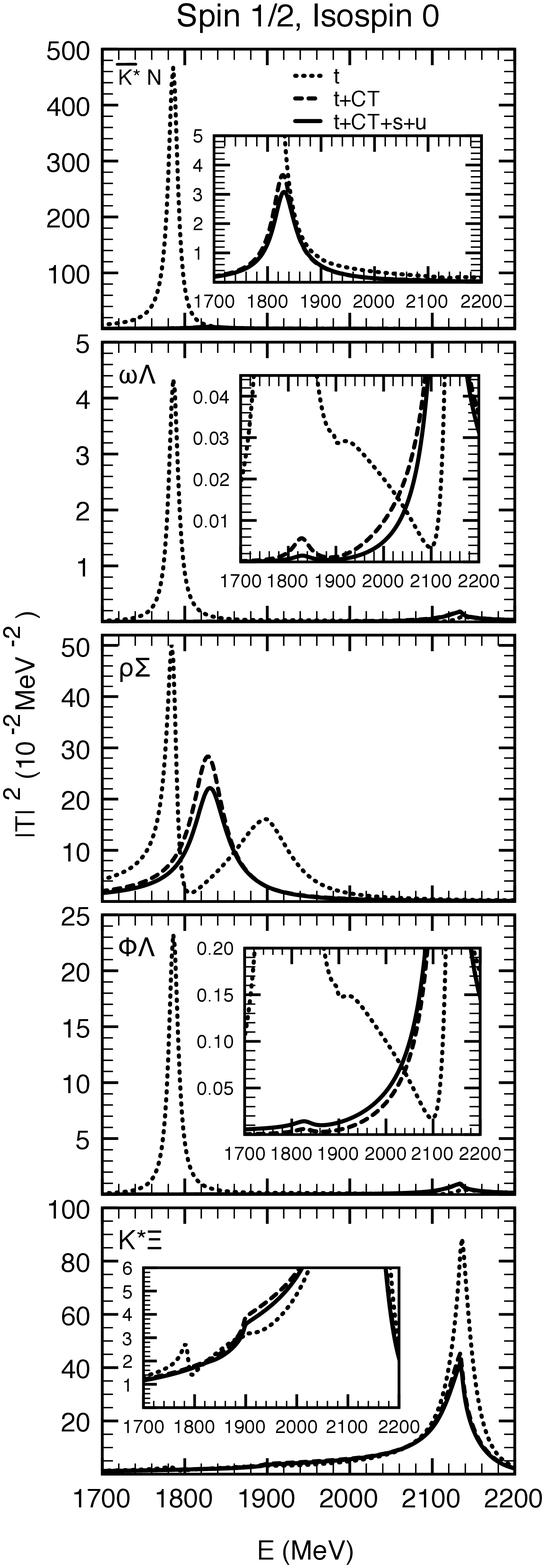}
\includegraphics[width= 0.38\textwidth,height=18cm]{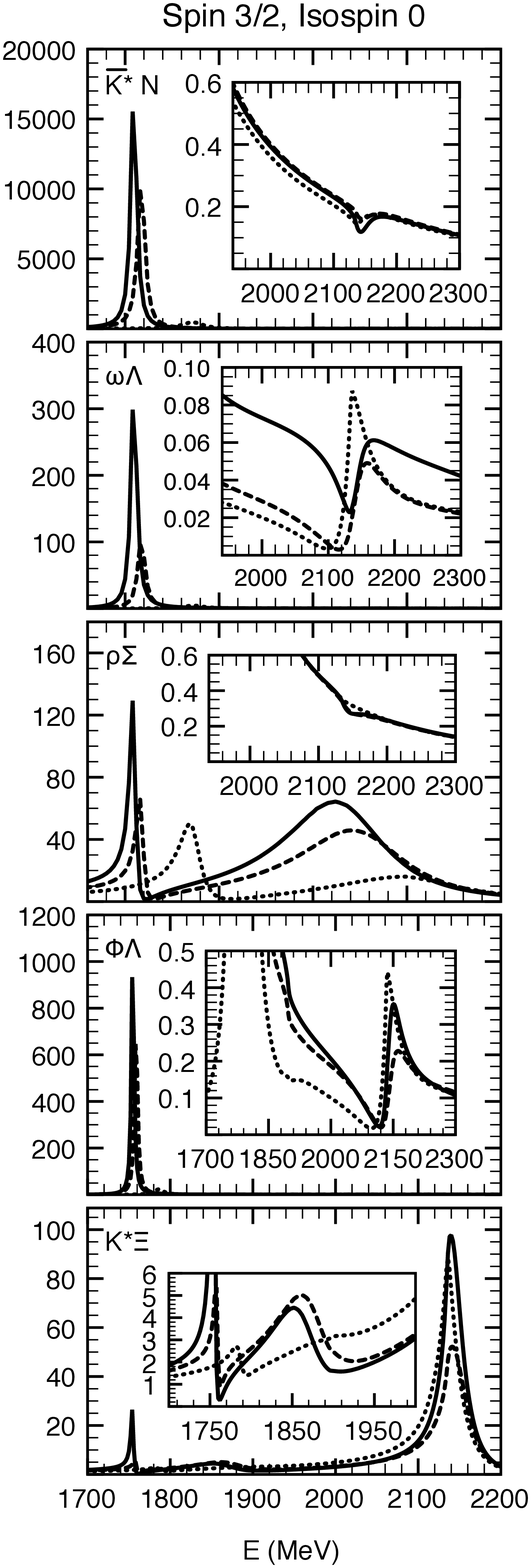}
\caption{Squared amplitudes for VB channels with total strangeness $-1$ and isospin 0. The right (left) panel shows the results  for VB systems in the spin 1/2 (3/2) configuration. }\label{fig_VBi0}
\end{figure}

The  total isospin of a VB system can be 0 or 1 and since we are looking for dynamically generated resonances, it is most relevant to study the low energy meson-baryon scattering considering the relative $s$-wave interaction. This implies that the total spin-parity of the VB systems in our work can be either $1/2^-$ or $3/2^-$. We show the VB squared amplitudes for isospin 0   in Fig.~\ref{fig_VBi0} for  spin 1/2 (right panel) as well as 3/2 (left panel). 
This figure depicts the amplitudes obtained by taking the $t$-channel exchange as the  VB interaction  (dotted lines), by adding the contact term given by Eq.~(\ref{contactv}) (dashed lines) and  the contributions from the $s$- and $u$- channels (solid line). The $t$-channel amplitudes are identical for both spins (although the scales of the figures on the right and left panels are different, which is due to the difference in the results obtained by adding other diagrams), and  show peaks corresponding to three poles (already discussed in Refs.~\cite{eulogiovb,us2}):  $1795 - i0$ MeV, $1923 - i4$ MeV and  $2138 - i21$ MeV.  However, by adding the contact interaction we find that none of the spin 1/2 poles survive as  physical poles (meaning that the poles  cannot be related to any physical state since they move in the complex plane to either become a virtual or a shadow pole), contrary to the spin 3/2 case where  three poles are found (see Tables~\ref{VB0coupling_shalf} and \ref
{VB0coupling_s3half}).
It is important to mention here that the poles listed in Tables~\ref{VB0coupling_shalf} and \ref{VB0coupling_s3half} are rather narrow because they have been obtained by calculating the loops without convoluting them over the large widths of the vector mesons, although we do follow this convolution procedure to calculate the amplitudes on the real axis. 
{\squeezetable
\begin{table}[t!]
\caption[]{Poles found in the  VB channels and their respective couplings for the isospin $0$, spin 1/2 case, when the interaction kernel is taken as the  $t$-channel diagram and when the contributions from the contact term, $s$- and $u$-channels are also added (i.e., $V = V_{sum} = V_{t} + V_{CT} + V_{s} + V_{u}$). Note that the superscript  $**$ indicates the unphysical nature of the pole (which means a virtual or a shadow pole).}  \label{VB0coupling_shalf}
\begin{ruledtabular}
\begin{tabular}{lrcrcrc}
&\multicolumn{2}{c}{$\xleftarrow{\hspace*{1cm}}$ Pole I$\xrightarrow{\hspace*{1cm}}$}&\multicolumn{2}{c}{ $\xleftarrow{\hspace*{1cm}}$Pole II$\xrightarrow{\hspace*{1cm}}$}&\multicolumn{2}{c}{ $\xleftarrow{\hspace*{1cm}}$Pole III$\xrightarrow{\hspace*{1cm}}$}\\
&\multicolumn{1}{c}{$V=V_t$}&\multicolumn{1}{c}{$V=V_{sum}$}&\multicolumn{1}{c}{$V=V_t$}&\multicolumn{1}{c}{$V=V_{sum}$}&\multicolumn{1}{c}{$V=V_t$}&\multicolumn{1}{c}{$V=V_{sum}$}\\
\hline
$M_R - i\Gamma/2$ (MeV) $\longrightarrow$ &$1795- i0$ &$1827 - i0^{\,**}$&$1923- i4$&$2044 - i61^{\,**}$&$2138- i21$&$2129 - i81^{\,**}$\\
\hline
Channels $\downarrow$& \multicolumn{6}{c}{Couplings ($g^i$) of the poles to the different channels} \\
\hline
$\bar{K}^*N$ $(1831)$             &$~~3.8 - i 0.0$  &-  &$~~0.1 - i 0.5$  &- &$~~0.0 - i 0.4$      &-\\
$\omega \Lambda$ $(1898)$ &$~~1.2 - i 0.0$  &-  &$~~0.3 - i 0.2$  &- &$-0.5 + i 0.3$     &-\\
$\rho \Sigma$ $(1963)$           &$-1.9 - i 0.0$     &-  &$~~3.7 + i 0.2$ &- &$~~0.1 + i 0.1$  &-\\
$\phi \Lambda$ $(2136)$        &$-1.8 - i 0.0$     &-  &$-0.5 +i 0.3$     &-  &$0.7 -i 0.4$        &-\\
$K^*\Xi$ $(2210)$                     &$-0.6 - i 0.0$     &-  &$~~1.0 + i 0.0$ &- &$~~4.2 + i 0.2$  &-
\end{tabular}
\end{ruledtabular}
\end{table}}

 {\squeezetable
\begin{table}[b!]
\caption[]{ Same as Table~\ref{VB0coupling_shalf} but for the spin 3/2 configuration.}\label{VB0coupling_s3half}
 \begin{ruledtabular}
\begin{tabular}{lrcrcrc}
&\multicolumn{2}{c}{$\xleftarrow{\hspace*{1cm}}$ Pole I$\xrightarrow{\hspace*{1cm}}$}&\multicolumn{2}{c}{ $\xleftarrow{\hspace*{1cm}}$Pole II$\xrightarrow{\hspace*{1cm}}$}&\multicolumn{2}{c}{ $\xleftarrow{\hspace*{1cm}}$Pole III$\xrightarrow{\hspace*{1cm}}$}\\
&\multicolumn{1}{c}{$V=V_t$}&\multicolumn{1}{c}{$V=V_{sum}$}&\multicolumn{1}{c}{$V=V_t$}&\multicolumn{1}{c}{$V=V_{sum}$}&\multicolumn{1}{c}{$V=V_t$}&\multicolumn{1}{c}{$V=V_{sum}$}\\
\hline
$M_R - i\Gamma/2$ (MeV) $\longrightarrow$ &$1795- i0$&$1760 - i0$&$1923- i4$&$1893 - i1$& $2138- i21$ &$2149 - i17$\\
\hline
Channels $\downarrow$& \multicolumn{6}{c}{Couplings ($g^i$) of the poles to the different channels} \\
\hline
$\bar{K}^*N$   $(1831)$            &$~~3.8 - i 0.0$ &$~~5.1 - i 0.0$    &$~~0.1 - i 0.5$  &$~~0.1 - i 0.2$    &$~~0.0 - i 0.4$       &$-0.1 - i 0.5$\\
$\omega \Lambda$ $(1898)$  &$~~1.2 - i 0.0$ &$~~1.9 - i 0.0$    &$~~0.3 - i 0.2$  &$~~0.4 - i 0.1$    &$-0.5 + i 0.3$      &$-0.3 + i 0.3$\\
$\rho \Sigma$    $(1963)$         &$-1.9 - i 0.0$    &$-1.3 + i 0.0$        &$~~3.7 + i 0.2$ &$~~4.7 + i 0.1$   &$~~0.1 + i 0.1$  &$~~0.1 + i 0.3$\\
$\phi \Lambda$  $(2136)$        &$-1.8 - i 0.0$    &$-2.5 + i 0.0$         &$-0.5 + i 0.3$    &$-0.2+i 0.1$         &$~~0.7 - i 0.4$   &$~~0.8-i 0.4$\\
$K^*\Xi$      $(2210)$                 &$-0.6 - i 0.0$    &$-0.9 + i 0.0$        &$~~1.0 + i 0.0$ &$~~1.9 + i 0.1$   &$~~4.2 + i 0.2$  &$~~4.5 + i 0.3$
\end{tabular}
\end{ruledtabular}
\end{table}}

\begin{figure}[b!]
\includegraphics[width= 0.38\textwidth,height=19cm]{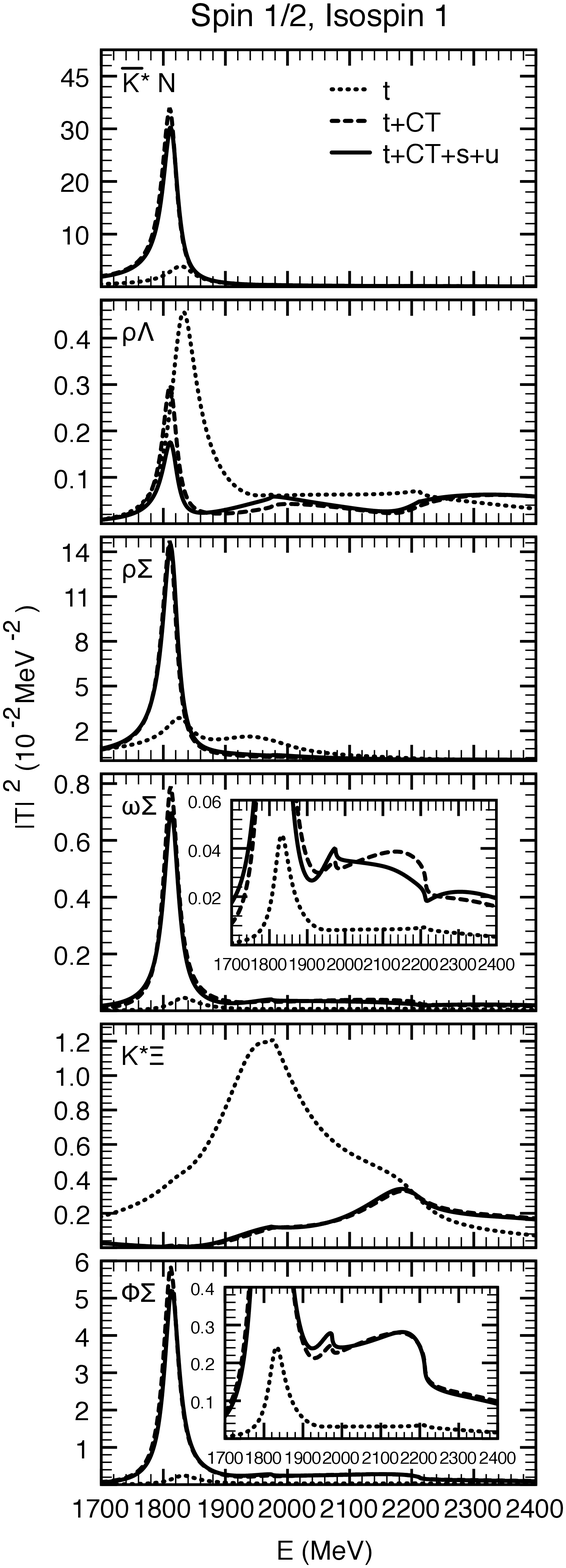}
\includegraphics[width= 0.38\textwidth,height=19cm]{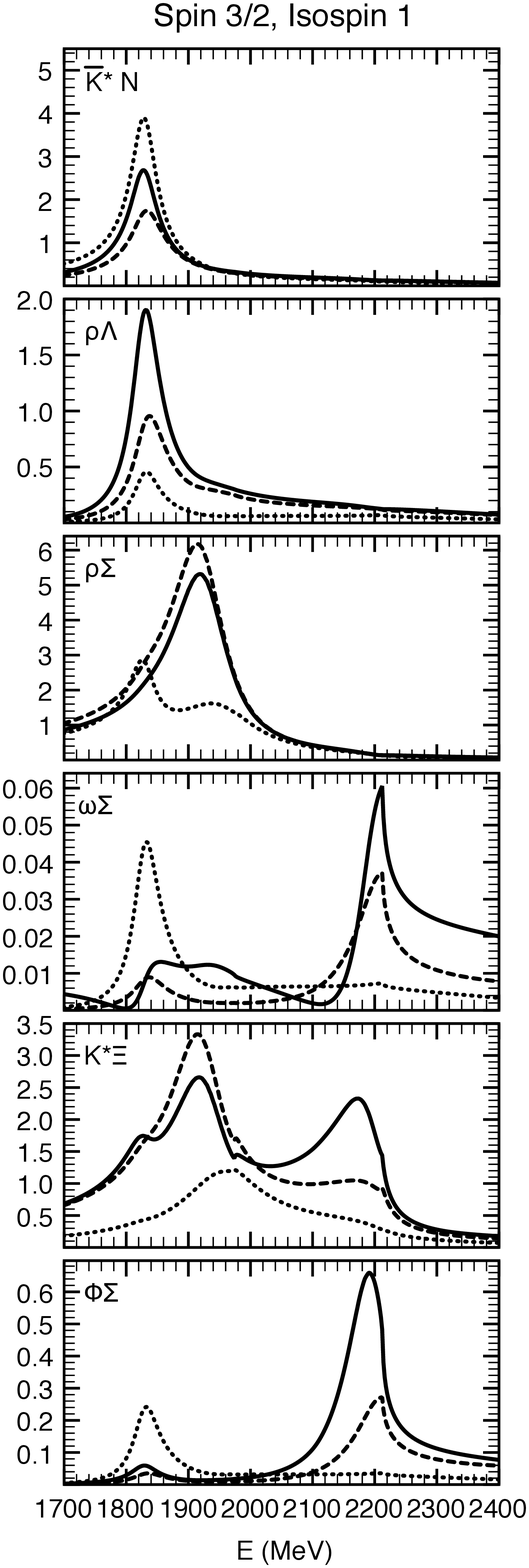}
\caption{Same as Fig.~\ref{fig_VBi0} but for the isospin 1 case.}\label{fig_VBi1}
\end{figure}
Next, let us discuss the results obtained for total isospin 1 VB systems. We show the squared amplitude for this case in Fig.~\ref{fig_VBi1}, where the different lines shown have the same meaning  as that in Fig.~\ref{fig_VBi0}. 
The amplitudes obtained with the $t$-channel interaction in the isospin 1 case show peaks around 1830 MeV and 1970 MeV, however none of them correspond to a pole in the complex plane and are cusp structures present near the thresholds of the $\bar{K}^* N$ and $\rho \Sigma$ channels, respectively. It should be added here that these cusp structures look like Breit-Wigner peaks due to the loops calculated with the convolution procedure. By adding the contact interaction and $s$- and $u$- channel diagrams, we find that the peak near 1830 MeV shifts 
to a lower energy by about 20 MeV in the spin 1/2 case  and the peak near the $\rho \Sigma$ threshold also shifts to a lower energy by 20 MeV in the spin 3/2 case and a corresponding resonance/bound state pole is found in both cases. The former  peak corresponds to a spin 1/2  bound state pole in the complex plane at $1822 - i0$  MeV and the latter one to a spin 3/2  resonance pole at 
$1947 - i5$ MeV (see Table~\ref{VB1coupling}).  Notice that a peak structure  can be seen  near 2.2 GeV in some spin 1/2 and 3/2 amplitudes (solid lines), which is just the $K^* \Xi$ cusp and does not signify the presence of any physical states. Additionally, the $\bar{K}^* N$ cusp continues to  show-up in the final spin 3/2 amplitudes (solid lines).
{\squeezetable
\begin{table}[htbp]
\caption{Poles and their couplings  to the VB channels in isospin 1 configuration. Like in Tables~\ref{VB0coupling_shalf}, \ref{VB0coupling_s3half}, here also, the results under the heading $V_{t}$ are those which have been obtained by taking the $t$-channel diagrams as the interaction kernels and those labelled with $V = V_{sum}$ (with $V_{sum} = V_{t} + V_{CT} + V_{s} + V_{u}$)  are the results of adding the contributions from the contact term, $s$- and $u$-channels. }  \label{VB1coupling}
\begin{ruledtabular}
\begin{tabular}{lrrcr}
Spin & \multicolumn{2}{c}{$\xleftarrow{\hspace*{1.5cm}} s  = 1/2 \xrightarrow{\hspace*{1.5cm}}$}&\multicolumn{2}{c}{$\xleftarrow{\hspace*{1.5cm}} s = 3/2 \xrightarrow{\hspace*{1.5cm}}$} \\
&\multicolumn{1}{r}{$V=V_t$}&\multicolumn{1}{r}{$V=V_{sum}$}&\multicolumn{1}{c}{$V=V_t$}&\multicolumn{1}{r}{$V=V_{sum}$}\\
\hline
$M_R - i\Gamma/2$ (MeV) $\longrightarrow$ &$-$ &$1822 - i0$&$-$&$1947 - i5$\\
\hline
Channels $\downarrow$& \multicolumn{4}{c}{Couplings ($g^i$) of the poles to the different channels} \\
\hline
$\bar{K}^*N$ $(1831)$              &$-$     &$~~2.3-i0.0$  &$-$  &$-0.3+i0.2$\\
$\rho \Lambda$ $(1886)$         &$-$    &$-0.6+i0.0$    &$-$  &$-0.5+i0.2$\\
$\rho \Sigma$ $(1963)$            &$-$     &$-1.9+i0.0$    &$-$  &$~~2.7+i0.2$\\
$\omega \Sigma$ $(1975)$      &$-$    &$-1.0+i0.0$    &$-$  &$~~0.3+i0.1$\\
$K^*\Xi$ $(2210)$                      &$-$    &$~~0.1-i0.0$  &$-$  &$~~1.9+i0.2$\\
$\phi \Sigma$ $(2213)$             &$-$    &$~~1.6-i0.0$  &$-$  &$~~0.2-i0.0$
\end{tabular}
\end{ruledtabular}
\end{table}}

We would like to postpone relating the $1/2^-$ poles found in our work to the known resonance states to the subsequent sections, where the coupling of VB and PB channels will be discussed. However,  the $3/2^-$ poles in our formalism do not couple to the PB channels in the s-wave, thus we can try to associate them to the known $\Lambda^* / \Sigma^*$ states.

 In case of $\Lambda^*$, there is only one known  resonance with $ J^\pi = 3/2^-$ in the 1700-2400 MeV energy region \cite{pdg}: $\Lambda~(2325)~D_{13}$.  None of the three $3/2^-$ states found in the present work seem to be compatible with this resonance. Thus our $3/2^-$ states, with the properties listed in Table~\ref{VB0coupling_s3half} are the predictions which should be tested in the experiments in the future. The evolution of these resonances in the VB systems, although non-trivial, is plausible since weakly bound molecule like states can exist near the threshold of the attractively interacting hadrons. 
  
 In the isospin one case also, there is only one $\Sigma $ resonance \cite{pdg} known to have quantum numbers $J^\pi = 3/2^-$ in the 1700-2400 MeV range:  $\Sigma~(1940)~D_{13}$. The state with isospin 1 and the spin-parity $3/2^-$ listed in Table~\ref{VB1coupling} is in good coincidence with the $\Sigma~(1940)~D_{13} $,  which has a significantly large branching ratio to the $\bar{K}^* N$ channel. It is also important to add that the full width at the half maximum of this state, from the calculations on the real axis is $\sim$  33 MeV, which is  still less than the known width of this state. It is possible that the $3/2^-$ states found in our work also couple to one/two pseudoscalar-octet/decuplet baryon channels, which are open at these energies and considering such additional channels can  increase the phase space for decay of this resonance. A coupling of $3/2^-$ states to the pseudoscalar-octet baryon channels can be made following the formalism of Ref.~\cite{javi}. Such improvements should be made in future.

\subsection{$\Lambda^*$ and $\Sigma^*$ in coupled PB-VB systems}
We have discussed the interaction of vector mesons and baryons and the resonances found in these systems so far. However, the pseudoscalar meson-baryon systems can also give rise to states with the same quantum numbers as the VB systems, and
some PB and VB channels have similar masses, thus treating them as  coupled channels can play an important role in understanding the properties of some of the resonances (as shown in Ref.~\cite{us2} for the low-lying resonances). Let us see now  how  the resonances found  in the VB systems, discussed in the  previous subsection, change when they are coupled to pseudoscalar mesons. As explained in Section~\ref{intII}, the PB-VB channels couple through a contact interaction obtained from an extension of the Kroll-Ruderman theorem by replacing the photons by the vector mesons. This s-wave interaction couples the VB and PB channels with total spin 1/2 only, thus the only the results with $J^\pi = 1/2^-$ discussed in the previous subsection get affected by the PB-VB coupling and the results for $J^\pi = 3/2^-$ remain unchanged. Indeed the calculations done within a different formalism in Ref.~\cite{javi} shows that the PB-VB coupling does very little to the $3/2^-$ VB channels and is more important in the $J^\pi = 1/2^-$ case. 

\begin{figure}[h!]
\includegraphics[width= 0.8\textwidth,height=13cm]{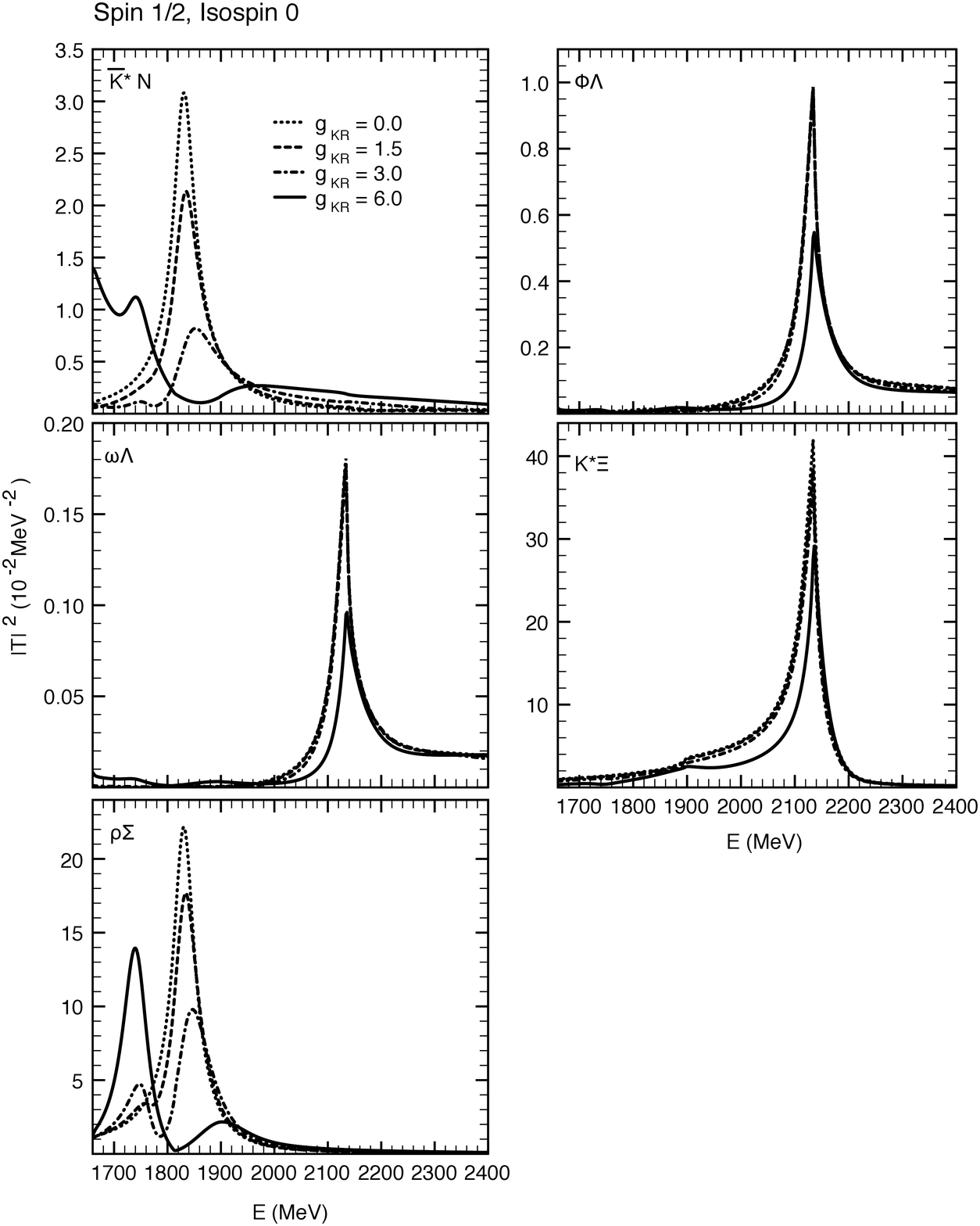}
\caption{Amplitudes of VB channels, in isospin 0, spin 1/2 configuration, when coupled to PB systems. The results are shown for different values of $g_{KR}$, which refers to the coupling between the PB and VB channels. From the KSRF relation (see Section~\ref{intII}) $g_{KR} = 6$.}\label{fig_pbvbi0_1}
\end{figure}

The amplitudes for the VB systems  with isospin 0, coupled to PB channels  are shown in Fig.~\ref{fig_pbvbi0_1}, where the results corresponding to uncoupled PB-VB systems are shown by a dotted line and marked by $g_{KR} = 0$. These results, for  
$g_{KR} = 0$, are identical to those shown with solid lines in Fig.~\ref{fig_VBi0} and as discussed in the previous subsection, the peaks appearing in these results  do not correspond to a pole in the complex plane which can be related to a physical state. We find, however, that the $\bar{K}^*N$ virtual pole given in Table~\ref{VB0coupling_shalf} turns into a resonant pole for $g_{KR} = 3$ coupling between the PB and VB channels (see Table~\ref{PBVB_coupling1}).
{\squeezetable
\begin{table}[h!]
\caption[]{ $g^{i}$ couplings of the spin $1/2$, isospin 0 poles to the VB and PB channels for the different strengths of  the  coupling of the Kroll-Ruderman term, $g_{KR}$ (PB-VB coupling). 
Note that no poles are found in isospin 0, spin 1/2 VB systems uncoupled to PB channels  (as shown in Table~\ref{VB0coupling_shalf}). However, one of the unphysical poles listed in Table~\ref{VB0coupling_shalf}: $1827 - i 0$ MeV, which is a $\bar{K}^* N$ virtual pole, turns into a resonance pole when PB channels are coupled to it.  This pole continues to be virtual for $g_{KR} = 1.5$ (the subscript $**$ indicates its virtual nature).} \label{PBVB_coupling1}
\begin{ruledtabular}
\begin{tabular}{ccrr}
Kroll-Ruderman &&&\\
coupling:  $g_{KR}$&\multicolumn{1}{c}{1.5}&\multicolumn{1}{r}{3}&\multicolumn{1}{c}{6}\\
\hline
$M_R - i\Gamma/2$ (MeV) $\longrightarrow$ &$1827. - i64.0^{**}$&$1859. - i62.0$&$1929 - i48$\\
\hline
Channels $\downarrow$& \multicolumn{2}{c}{Couplings ($g^i$) of the poles to the different channels} \\
\hline
$\bar{K}N$ $(1435)$                      &-&$ - 0.2 + i 0.6$&$ -0.2+i0.4$\\
$\pi \Sigma$ $(1330)$                   &-&$ -0.1 + i 0.2$& $ -0.2-i0.1$\\
$\eta \Lambda$ $(1663)$             &-&$ -0.1 + i 0.4$& $ 0.0+i0.4$ \\
$K\Xi$ $(1814)$                              &-&$0.2 - i 0.3$& $ 0.3-i0.5$\\ 
$\bar{K}^*N$ $(1831)$                   &-&$-2.0 + i 1.1$&$ -1.1+i0.9$\\
$\omega \Lambda$ $(1898)$        &-&$-0.1 + i 0.1$& $ 0.3+i0.2$\\
$\rho \Sigma$ $(1963)$                  &-&$ 6.0 - i 0.3$& $ 3.4+i1.1$\\
$\phi \Lambda$ $(2136)$               &-&$ 0.1 - i 0.1$& $ -0.5-i0.3$\\
$K^*\Xi$ $(2210)$                           &-&$-0.6 + i 0.3$&$ -1.7+i0.3$\\
\end{tabular}
\end{ruledtabular}
\end{table}}
 It moves closer to the real axis on increasing the coupling to 6 and ends up at $1929 - i48$ MeV. This pole 
seems to couple strongly to the $\rho \Sigma$ channel and might be related to the poorly understood $\Lambda (2000)$ with unknown spin-parity. The mass and width of the pole found in our work, $1929 - i48$ MeV, are in good agreement with those known for the $\Lambda (2000)$. Further, this resonance has been found in the reactions with the $\bar{K}^* N$ final state, which is an open channel with  the largest coupling strength (as shown in Table~\ref{PBVB_coupling1}) .

Before going ahead we would like to briefly discuss about the existence of an ambiguity of the sign in the
relative phase of the couplings of the  PB and VB systems to a resonance. The source of the ambiguity lies in the fact that the PB $\leftrightarrow$ VB transition amplitudes
coming from the Lagrangian given by Eq.~(\ref{pbvb})
are such that 
$T_{\rm{PB} \rightarrow \rm{VB}}=-T_{{\rm VB}\rightarrow {\rm PB}}$ (which in turn is the result of the purely imaginary nature of the kernel $V_{\rm{PB} \rightarrow \rm{VB}}$).
Thus, to calculate the couplings we need to fix our convention. We do so by writing the amplitudes in terms of  the effective couplings of a
resonance to its constituent PB and VB hadron-pairs 
as
\begin{eqnarray}
 T_{{\rm PB} \rightarrow {\rm P'B'}}&\equiv& g_{{\rm PB}}\frac{1}{\sqrt{s}-M_R+i\Gamma/2}g_{\rm P'B'}, \label{eq:Tii}\\
 T_{{\rm VB} \rightarrow {\rm V'B'}}&\equiv& \left(i g_{{\rm VB}}\right)\frac{1}{\sqrt{s}-M_R+i\Gamma/2}\left(-ig_{\rm V'B'}\right) ,
\label{eq:Tjj}
\end{eqnarray}
and  we  write the non-diagonal amplitudes consistently as 
\begin{eqnarray}
 T_{{\rm PB} \rightarrow {\rm VB}}&\equiv& g_{{\rm PB}}\frac{1}{\sqrt{s}-M_R+i\Gamma/2} \left( - i g_{\rm VB} \right), \label{eq:Tij}\\
 T_{{\rm VB} \rightarrow {\rm PB}}&\equiv& \left(i g_{{\rm VB}}\right)\frac{1}{\sqrt{s}-M_R+i\Gamma/2}ig_{\rm PB} =   - T_{{\rm PB} \rightarrow {\rm VB}}
\label{eq:Tji}
\end{eqnarray}
Formally, Eq.~(\ref{eq:Tji}) is obtained by taking the complex conjugate of Eq.~(\ref{eq:Tij}) with the phase of the couplings $g_{\rm PB}$ and $g_{\rm VB}$ kept unchanged. However, equivalent amplitudes can also be obtained from effective Lagrangians written in terms of an ``effective field"  of the resonance and Eqs.~(\ref{eq:Tii}-\ref{eq:Tji}) have been written  with this prescription, which has been used to calculate the couplings of the resonance to different meson-baryon channels.

\begin{figure}[ht!]
\includegraphics[width= 0.9\textwidth,height=12cm]{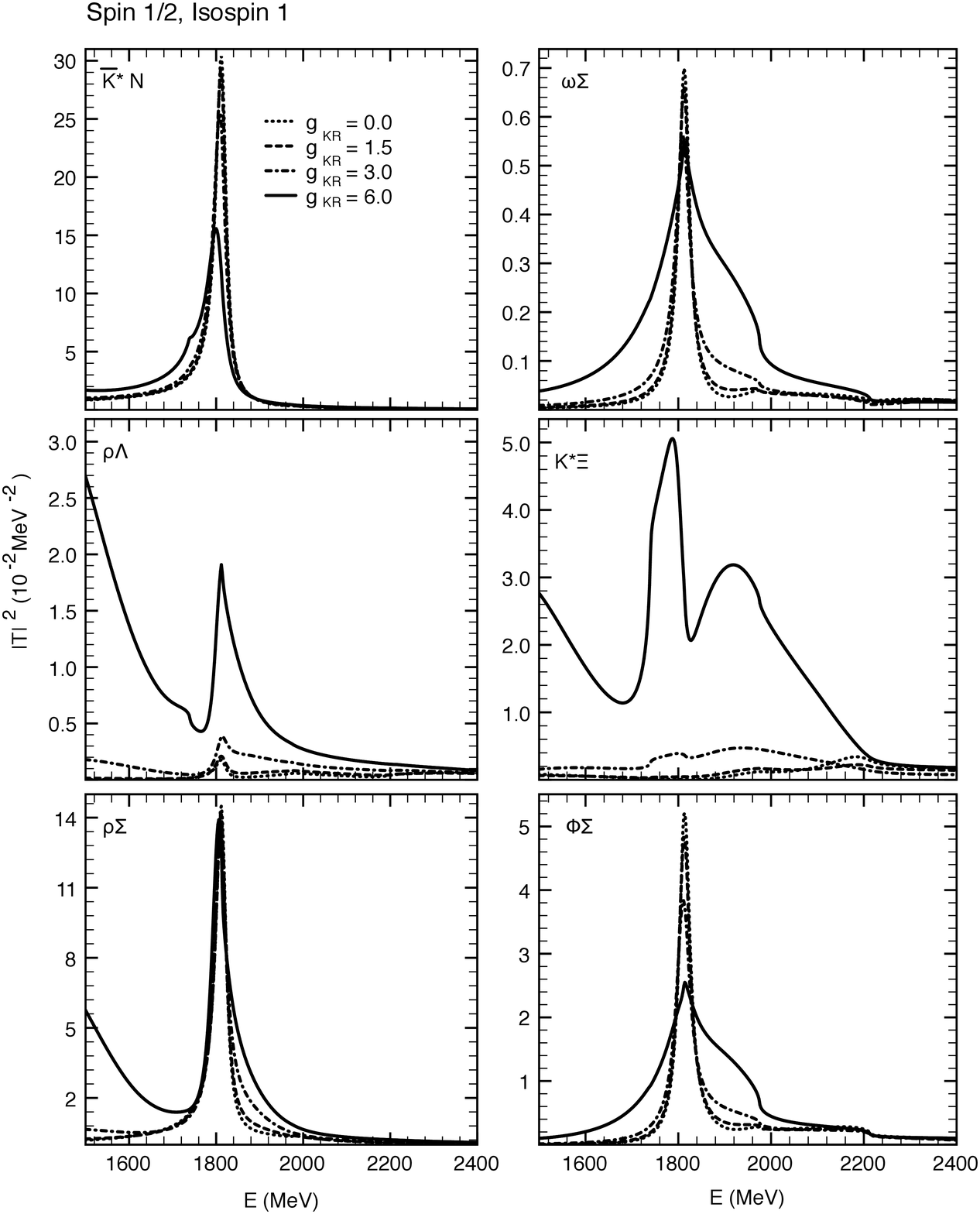}
\caption{Same as Fig.~\ref{fig_pbvbi0_1} but for the total isospin 1.}\label{fig_pbvbi1_1}
\end{figure}

Next, we discuss the results for isospin 1, spin 1/2 PB-VB coupled channels. The squared amplitudes obtained in this case are shown in Fig.~\ref{fig_pbvbi1_1}. 
As shown in Table~\ref{VB1coupling},  a pole with mass $1822 - i0$ MeV was found  in the VB systems in isospin 1, spin 1/2 configuration. The dotted lines in Fig.~\ref{fig_pbvbi1_1} show the corresponding peak in the squared amplitudes calculated on the real axis. We find that coupling this pole to the pseudoscalar mesons  does not alter its properties much and results in only  a little increase in its width.  
This  can be understood by looking at the weak coupling of this pole to the different PB channels given in Table~\ref{table_pbvb_iso1}. 
{\squeezetable
\begin{table}[h!]
\caption{ $g^{i}$ couplings of the spin $1/2$, isospin 1 poles to the VB and PB channels for different strengths of  the coupling of the Kroll-Ruderman term. Note that three poles appear for PB-VB coupling $g_{KR} = 6$. }\label{table_pbvb_iso1}
\begin{ruledtabular}
\begin{tabular}{lrrrrr}
Kroll-Ruderman coupling:  $g_{KR}$&\multicolumn{1}{c}{0}&\multicolumn{1}{c}{3}&\multicolumn{3}{c}{ $\xleftarrow{\hspace*{2cm}}$ 6 $\xrightarrow{\hspace*{2cm}}$ }\\
\hline
$M_R - i\Gamma/2$ (MeV) $\longrightarrow$ &$1822- i0$ &$1821- i1$&$1821-i16$& $1873 - i88$&$1936 - i132$\\
\hline
Channels $\downarrow$& \multicolumn{5}{c}{Couplings ($g^i$) of the poles to the different channels} \\
\hline
$\bar{K}N$ $(1435)$                      &$0.0-i0.0$&$0.1+i0.1$&$ 0.1+i0.4$&$ 0.9-i0.6$& $ -0.2-i0.4$\\
$\pi \Sigma$ $(1330)$                   &$0.0-i0.0$&$0.0-i0.0$& $-0.1+i0.1$&$ 0.8-i0.3$&$ -0.1-i0.5$\\
$\pi \Lambda$ $(1253)$                &$0.0-i0.0$&$0.2-i0.1$& $ 0.5-i0.1$& $ 0.2-i0.6$& $ -0.2+i0.8$ \\
$\eta \Sigma$ $(1740)$                 &$0.0-i0.0$&$-0.3-i0.2$& $ -0.5-i0.6$ &$ -0.4+i0.7$&$ 0.9-i1.0$\\
$K\Xi$ $(1814)$                              &$0.0-i0.0$&$0.2+i0.0$& $ 0.5+i0.2$&$ -1.0-0.9$&$ -0.4+i1.1$\\ 
$\bar{K}^*N$ $(1831)$                   &$2.3-i0.0$&$2.5+i0.3$& $ 2.7+i1.2$& $ -0.8-i0.1$&$ 0.9+i0.1$\\
$\rho \Lambda$ $(1886)$               &$-0.6+i0.0$&$-0.6+i0.1$& $ -1.2+i0.3$&$ 2.8+i2.4$&$ -0.4-i0.2$\\
$\rho \Sigma$ $(1963)$                  &$-1.9+i0.0$&$-2.0+i0.2$& $ -2.7+i0.4$&$ 4.3+i1.9$& $ -2.7-i3.4$\\
$\omega \Sigma$ $(1975)$            &$-1.0+i0.0$&$-0.9-i0.0$& $ -0.5+i0.2$&$ -0.1-i1.6$&$ -2.4-i0.3$\\
$K^*\Xi$ $(2210)$                           &$0.1-i0.0$&$-0.1-i0.3$& $ -0.3-i1.9$&$ -3.0+i1.3$& $ 4.7+i0.5$\\
$\phi \Sigma$ $(2213)$                   &$1.6-i0.0$&$1.5+i0.1$& $ 1.0.-i0.2$&$ -0.0+i2.0$&$3.5+i0.7$\\
\end{tabular}
\end{ruledtabular}
\end{table}}
Thus, even though the phase space for this state increases by coupling lighter decay channels to it, its weak coupling to the latter does not lead to a larger width. Interestingly, we find that two more isospin 1, spin 1/2 poles emerge when PB and VB channels are coupled: $1873 - i88$ MeV and $1936 - i132$ MeV. These poles are found for $g_{KR} = 6$ and possess large couplings with the $\rho \Sigma $ and $K^* \Xi$ channnels.

Let us now check if the isospin 1 states found in our work can be related to any of the known resonances of Ref.~\cite{pdg}. We find the pole $1821-i16$ MeV to be compatible with the $\Sigma (1750) S_{11}$. It is important to mention that the Breit-Wigner mass and width of this state extracted from the amplitude on the real axis: $1800 - i28$ MeV are even more compatible with the properties of $\Sigma (1750) S_{11}$. These values found on the real axis are more compatible since they include the effect of the large decay widths of the vector mesons. Another  known feature of $\Sigma (1750) S_{11}$ is its  large branching ratio (15-55 $\%$)  to the $\eta \Sigma$ channel, which is also in agreement with our finding of its coupling to the $\eta \Sigma$ channel being the  largest amongst the open channels (Table~\ref{table_pbvb_iso1}).

The other two isospin 1 poles found here: $1873 - i88$ MeV and $1936 - i132$ MeV, seem to be compatible with the information listed for the $\Sigma (2000) S_{11}$ resonance with its mass ranging from 1755 MeV to 2000 MeV in different experiments.  Since the width of the two poles are large and the difference between their masses is relatively small, it leads to merging of the two possible corresponding peaks in the cross sections. In fact, it is hard to see a well distinguished peak corresponding to these poles in Fig.~\ref{fig_pbvbi1_1}, except in case of the $K^* \Xi$ channel (which is a closed channel for the decay of these resonances). Only an enhancement/a broad bump is seen in the 1850-2000 MeV energy region, in  the  squared amplitudes of all the channels shown in Fig.~\ref{fig_pbvbi1_1} (with solid lines). It is, thus, quite possible that $\Sigma (2000) S_{11}$ might be linked to two closely spaced poles, as  found in our work.

\begin{figure}[h!]
\includegraphics[width= 0.68\textwidth,height=11cm]{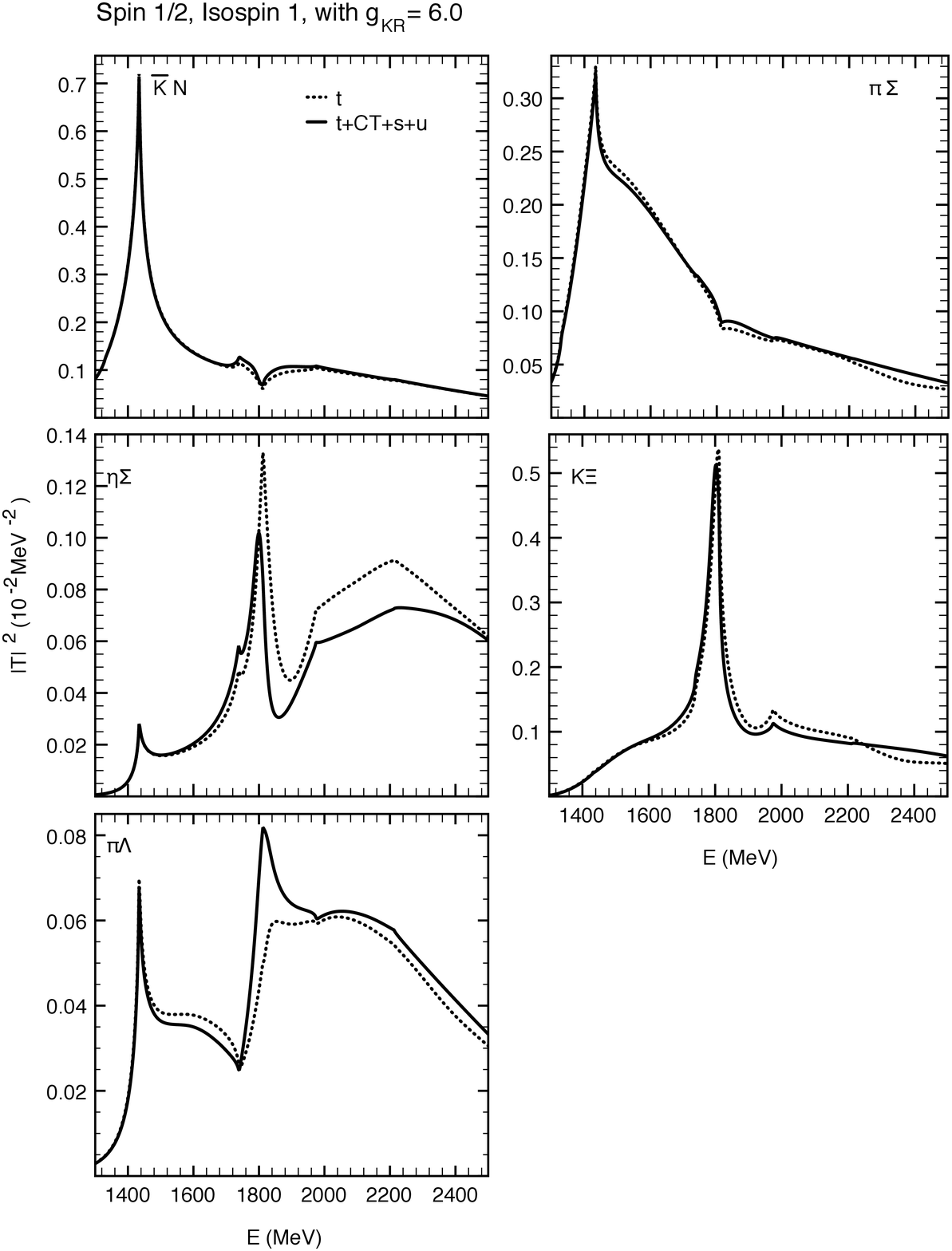}
\caption{Amplitudes for the pseudoscalar-baryon channels in isospin 1. The dotted (solid) line is the result of the calculation of the coupled channel scattering equations for the PB-VB systems by taking the 
VB  interaction obtained from the $t$-channel diagrams (by summing the $s$-, $t$-, $u$-channel diagrams and the contact term of Eq.~(\ref{contactv})). }\label{fig_pbvbi1_2}
\end{figure}

We show  the squared amplitudes for the PB  channels in Fig.~\ref{fig_pbvbi1_2}, 
coupled to VB, for the isospin 1 case. This figure shows a comparison of the results obtained for coupled PB-VB systems (with $g_{KR} = 6$) when the VB $\rightarrow$ VB amplitudes are obtained (1) from the $t$-channel interaction and (2) by summing the contributions from the $s$-, $t$-, $u$-channel diagrams and the contact term explained in Section~\ref{intI}
\begin{equation}\label{vsum}
 V_t + V_s + V_u + V_{CT}.
\end{equation}
It can be seen that the PB amplitudes at low energy do not get altered by the two different VB interactions, implying that one can rely on $t$-channel diagrams (for both PB and VB channels) while studying the low lying PB resonances. In fact, the PB amplitudes at higher energies, where VB channels are open, also do not change from one case to the other except for the appearance of a clearer signal of a resonance at 1800 MeV in the $\eta \Sigma$ and $\pi \Lambda$ channels, when the calculations are done by using the VB interactions obtained from  Eq.~(\ref{vsum}). These findings can be easily understood by noticing  weaker couplings of the resonances, listed in Table~  \ref{table_pbvb_iso1}, to the PB channels as compared to those of the VB channels. This also implies that it is more advantageous to study the resonances in the 1700-2200 MeV region  by analyzing the reactions producing VB channels.

\begin{figure}[h!]
\includegraphics[width= 0.7\textwidth,height=9cm]{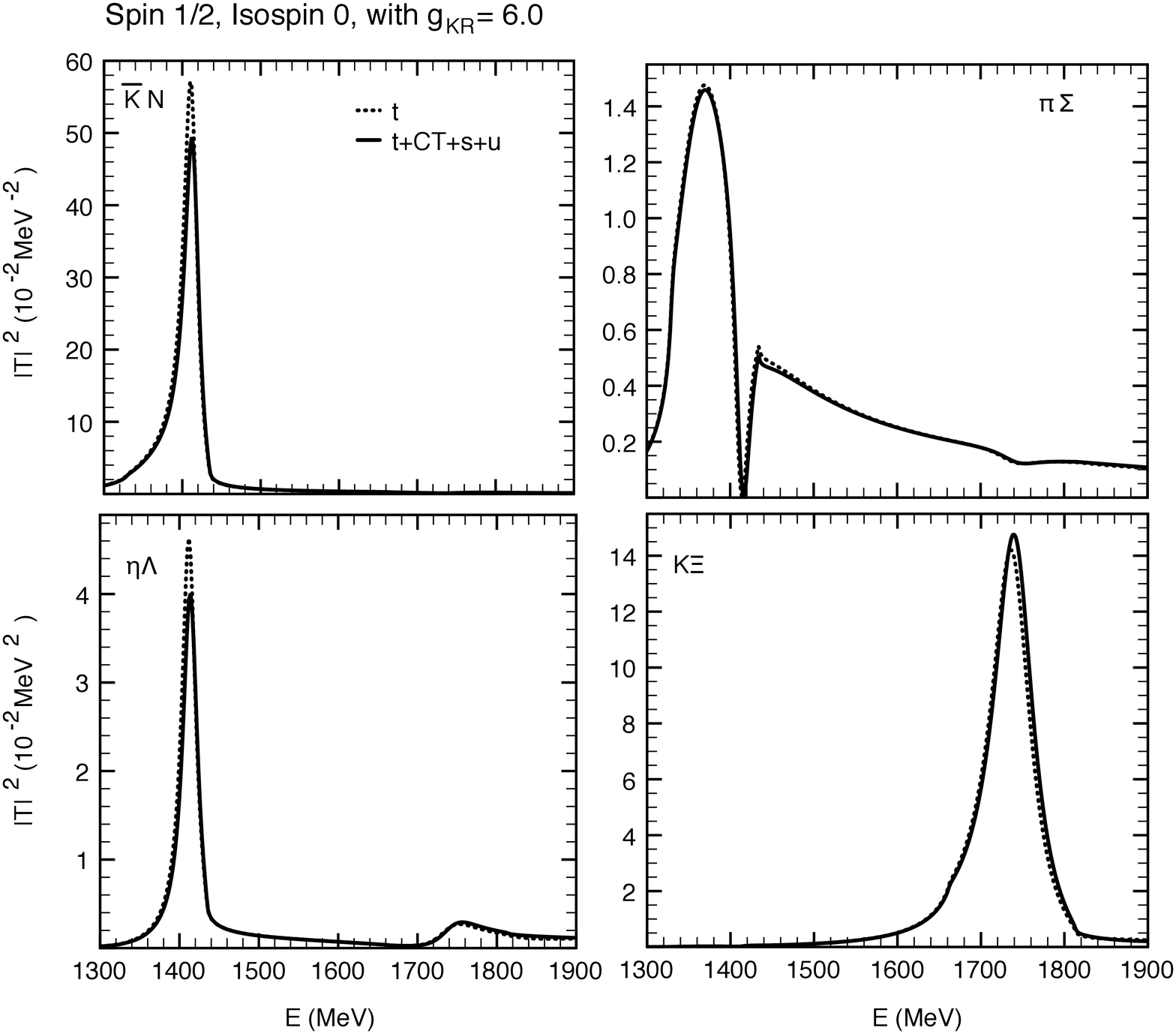}
\caption{Amplitudes for pseudoscalar-baryon channels in isospin 0. The lines here have the same meaning as in Fig.~\ref{fig_pbvbi1_2}. }\label{fig_pbvbi0_2}
\label{fig_pbVB1_i0}
\end{figure}
The amplitudes for PB channels in isospin 0 also do not alter much by taking different VB interactions (see Fig~.\ref{fig_pbvbi0_2}),
once again due to their weak coupling to the $\Lambda$ resonances generated in VB systems (see Table~\ref{PBVB_coupling1}), once again implying the importance of studying reactions with VB final states to identify the properties of the VB resonances. Finally, we do not discuss  the low-lying resonances in detail here since their properties  change very little by using the VB amplitudes of Eq.~(\ref{vsum}) in the calculations and hence the results obtained in Ref.~\cite{us2} do not change much. Even though the changes are little,  for the sake of completeness, we give the low-lying poles and their coupling to different PB-VB channels in Table~\ref{lowlying_poles}.

{\squeezetable
\begin{table}[htbp]
\caption{Couplings of the low lying $J^\pi = 1/2^-$ poles  to the PB and VB channels (see Ref.~\cite{error}).}  \label{lowlying_poles}
\begin{ruledtabular}
\begin{tabular}{lrrr|lrr}
\multicolumn{4}{c|}{$\xleftarrow{\hspace*{3cm}}$Isospin $0 \xrightarrow{\hspace*{3cm}}$}&\multicolumn{3}{c}{$\xleftarrow{\hspace*{1.5cm}}$Isospin 1$\xrightarrow{\hspace*{1.5cm}}$}\\
$M_R - i\Gamma/2$ (MeV) $\longrightarrow$ &$1358 - i53$ &$1414 - i12$ & $1746 - i28$&  &$1427 - i 145$  &$1438 - i  198$ \\
\hline
Channels $\downarrow$& \multicolumn{3}{c}{Couplings }&Channels$\downarrow$ & \multicolumn{2}{c}{Couplings } \\
\hline
$\bar{K} N$   $(1435) $            &$~1.2 - i 1.4$&$~2.8 + i 0.6$&$~0.3 - i 0.6$&$\bar{K} N$ $(1435) $               &$-0.8 + i 1.1$&   $ 0.4 - i 0.9$   \\
$\pi \Sigma$    $(1330)$          &$-2.2 + i 1.4$&$-0.2 - i 1.1$&$~0.1 + i 0.3$&$\pi \Sigma$  $(1330)$             &$ 1.9 - i 1.9$&  $ 1.7 - i 1.4$   \\
$\eta \Lambda$ $(1663)$       &$~0.1 - i 0.6$&$~1.5 + i 0.1$&$-1.0 + i 0.3$&$\pi \Lambda$ $(1253)$        &$ -0.3 - i 0.1$&  $-0.8 + i 1.2$    \\
$K \Xi$     $(1814)$                   &$-0.6 + i 0.4$&$~0.0 - i 0.3$&$~3.4 + i 0.3$&$\eta \Sigma$  $(1740)$         &$-0.2 - i 0.1$&$-0.5 + i 0.9$    \\
$\bar{K}^*N$ $(1831)$            &$ 0.4 + i 1.6$&$-4.9 - i 0.0$&$0.9 + i 0.3$&$K \Xi$     $(1814)$                    &$ 0.9 - i 0.8$& $ 1.3 - i 1.7$     \\
$\omega \Lambda$$(1898)$ &$-0.4 + i 0.7$&$-1.6 - i 0.2$&$-0.2 - i 0.2$&$\bar{K^*}N$ $(1831)$            &$-1.7 + i 0.0$&$-3.1 - i 0.1$\\
$\rho \Sigma$     $(1963)$      &$~6.4 - i 1.3$&$-1.5 + i 2.3$&$-3.2 - i 0.3$&$\rho  \Lambda$ $(1886)$       &$6.0 - i 0.3$&    $7.3 - i 0.9$   \\
$\phi \Lambda$  $(2136)$      &$~0.5 - i 1.0$&$2.3 + i 0.3$&$0.3 + i 0.2$&$\rho  \Sigma$  $(1963)$        &$-4.5 - i 0.1$&    $-8.0 + i 1.2$  \\
$K^*\Xi$       $(2210)$              &$~5.4 - i 1.3$&$-0.6 + i 1.9$&$-0.7 - i 0.5$&$\omega  \Sigma$   $(1975)$&$0.1 - i 0.4$&   $-0.3 - i 0.2$   \\
						&			&			&		  &$K^* \Xi$$(2210)$                     &$-6.1 + i 1.0$&    $-6.2 + i 1.1$  \\
						&			&			&	            &$\phi  \Sigma$  $(2213)$         &$-0.2 + i 0.5$&   $0.5 + i 0.3$  \\
\end{tabular}
\end{ruledtabular}
\end{table}}

\section{Summary}
We have studied the generation of baryon resonances in a wide energy range (1300 - 2200 MeV) in pseudoscalar-baryon and vector meson-baryon strangeness $-1$ coupled channels.
We find ten states in our work which are summarized in Table~\ref{summary}, together  with their quantum numbers and with the possible corresponding known states. In Table~\ref{summary}, we list the pole positions
found in the complex plane as well as the masses and widths deduced from the peaks found in the squared amplitudes calculated on the real axis. It is important to give
this latter information since the calculations are done in the complex plane assuming zero width for the $\rho$ and $K^*$ mesons due to the complications explained in
Refs.~\cite{eulogiovb,us2}. Thus, it is more meaningful to compare the masses and widths of the states obtained from the amplitudes calculated on the real axis with the properties of the known resonances.
However, we find it difficult to obtain this information clearly  in case of the isospin 1 poles found near 1430 MeV and hence we do not give it.  As can be seen in Table~\ref{summary}, apart from the relatively well known low-lying resonances, we find clear evidences for  $\Lambda (2000)$, $\Sigma(1750)$, $\Sigma(1940)$ and $\Sigma(2000)$, the last of which can be related to two poles.  Our work can be helpful in determining  the unknown spin-parity of $\Lambda(2000)$ to be $1/2^-$. In addition to this, we predict the existence of three $\Lambda^*$'s and a $\Sigma^*$.
{\squeezetable
\begin{table}[h]
\caption[]{A summary of strangeness $-1$ baryon  resonances found in the present work together with the corresponding known $\Lambda^*, \Sigma^*$'s. We list the pole positions found in the complex plane as well as the peak position and half width obtained from the amplitudes calculated on the real axis. The source of the difference between the two comes from the (non-) consideration of the width of the vector mesons in the (former) latter case.}  \label{summary}
\begin{ruledtabular}
\begin{tabular}{c|cccc}
&&\multicolumn{2}{c}{E $- i\Gamma/2$ (MeV)} &\\
& Spin-parity ($J^\pi$) &In the complex plane & On the real axis & Related known states\\
\hline
\multirow{6}{*}{Isospin 0}&$1/2^-$& $1358 - i53$, $1414 - i12$     &$1370 - i36$
, $1412 - i12$
&$\Lambda (1405)$\\
&$1/2^-$& $1746 - i28$                              & $1739 - i28$&$\Lambda (1670)$\\
&$3/2^-$& $1760 - i0$                                 &$1754 - i2$ & $\Lambda ?$ \\
&$3/2^-$& $1893 - i1$                                 &$1862 - i33$ & $\Lambda ?$\\
&$1/2^-$& $1929 - i48 $                             &$1903 - i60$ &$\Lambda (2000)$  \\
&$3/2^-$& $2149 - i17$                               & $2139 - i16$&$\Lambda ?$\\
\hline
\multirow{4}{*}{Isospin 1}&$1/2^-$& $1427 - i 145$, $1438 - i 198$& $...$& $\Sigma ?$\\
&$1/2^-$& $1821 -i16$&$1800 - i28$&$\Sigma (1750)$ \\
&$1/2^-$&$1873 - i88$, $1936 - i132$&$\sim 1920 - i150$& $\Sigma (2000)$\\
&$3/2^-$& $1947 - i5$&$1918 - i60$ &$\Sigma (1940)$ \\
\end{tabular}
\end{ruledtabular}
\end{table}}

Further, we find that  the VB interaction obtained from $t$-channel diagrams is reliable to study their couplings to the low-lying PB resonances. However,
in agreement with the results found in Ref.~\cite{us} for the strangeness 0 case, the other sources of diagrams discussed in Section~\ref{intI} play a very important 
role in understanding the properties of the VB resonances. Another finding of our work is that the pseudoscalar-baryon channels couple very weakly
to several resonances found to get generated in VB systems, which implies that it is not suitable to study these resonances by analyzing the reactions
with pseudoscalar meson-baryon final states. Rather, it is desired to consider the reactions producing vector mesons to scrutinize these resonances.

\section{Acknowledgements}
 This work is partly  supported  by the Grant-in-Aid for Scientific Research on Priority Areas titled ÒElucidation of New Hadrons with
a Variety of Flavors" (E01: 21105006 for K.P.K and A.H) and (22105510 for H.~N) and the authors acknowledge the same. A.~M.~T  
is thankful to the support from the Grant-in-Aid for the Global COE Program ÒThe Next
Generation of Physics, Spun from Universality and EmergenceÓ from the Ministry of Education,
Culture, Sports, Science and Technology (MEXT) of Japan.


\begin{thebibliography}{99}
\bibitem{error} 
Please notice that in some cases we have a phase difference of $i$ ( or $-i$) with respect to the corresponding couplings mentioned in 
 Ref.~\cite{us2} in order to be consistent with the definitions given by Eqs.~(\ref{eq:Tii}) and (\ref{eq:Tjj}).
\end{thebibliography}
\end{document}